\documentclass{aa}
\usepackage{graphicx}
\usepackage{txfonts}
\begin{document}
   \title{Do Black Holes End up as Quark Stars ?}

   \subtitle{}
  
   \author{R.K.Thakur
	\inst{}
	  }

   \offprints{R.K.Thakur}

   \institute{Retired Professor of Physics,
   School   of   Studies in Physics,
   Pt. Ravishankar Shukla University,
   Raipur\\
\email{rkthakur0516@yahoo.com}
    }

   \date{Received: ..........}

   \abstract{
The possibility of the existence of quark stars has been discussed by several 
 authors since 1970. Recently, it has been pointed out that two putative neutron stars,
 RXJ 1856.5 - 3754 in Corona Australis and 3C58 in Cassiopeia are too small and too
 dense to be neutron stars; they show evidence of being quark stars. Apart from these
 two objects, there are several other compact objects which fit neither in the 
 category of neutron stars nor in that of black holes. It has been suggested that
 they may be quark stars.In this paper it is shown that a black hole cannot collapse
 to a singularity, instead it may end up as a quark star. In this context
 it is shown that a gravitationally collapsing black hole acts as an 
 ultrahigh energy  particle accelerator, hitherto inconceivable in any 
 terrestrial laboratory, that  continually accelerates particles comprising 
 the matter in the black hole. When the energy \textit{E} of the particles in the
 black hole is $\geq 10^{2} \mathrm{GeV}$, 
 or equivalently the temperature   \textit{T} of the matter in the black holes 
 is $\geq 10^{15}\mathrm{K}$, the  entire matter in the black hole will be converted
 into quark-gluon plasma permeated  by leptons. Since quarks and leptons are spin 1/2
  particles,they are governed by  Pauli's exclusion principle. Consequently, one 
  of the two possibilities will occur; either Pauli's exclusion principle would 
  be violated and the black hole would collapse to a singularity, or the collapse of
  the black hole to a singularity would be inhibited by Pauli's exclusion principle, and
  the black hole would eventually explode with a mini bang of a sort. After explosion, the remnant 
   core would stabilize as a quark star. 

   \keywords{black hole, gravitational collapse, space-time singularity, quark star}
   }
\titlerunning{Do Black Holes End up as Quark Stars ?}
\authorrunning{R.K.Thakur}
   \maketitle
%
\section{Introduction} 
The possibility of the existence of the quark stars,i.e. the stars 
  composed of the fundamental constituents of matter, viz., quarks is being 
  discussed for more than three decades (Itoh 1970, Bodmer 1971, Collins and 
  Perry 1975,Brecher and Caporaso 1976, Chaplin and Nauenberg 1978,Witten
   1984, Alcock et al. 1986, Haensel et al. 1986, Li et al. 1995, Bombaci 1997, Cheng
    et al. 1998, Xu et al. 1999). It has been suggested that such an object
 would have an approximately thermal spectrum (Xu 2002, Pons et al. 2002).
  Moreover,  in this context it may be noted that there are several compact
   objects, e.g.,Her $-$ X1, 4U 1820 $-$ 30 (Bombaci 1997, Dey et. al. 1998),
   SAX J 1808.4 $-$ 3658 (Li et al.1999 a), 4U1728 $-$ 34 (Li et al. 1999 b), 
   PSR 0943+10 ( Xu et al. 1999), which fit neither in the category of
 neutron stars nor in that of black holes. The apparent compactness of 
 these objects could be explained if they are composed of quarks.
 
               Recently, two teams $-$ one led by David Helfand of Columbia 
 University, New York (Slane et al. 2002), and another led by Jeremy Drake 
 of Harvard-Smithsonian Center for  Astrophysics, Cambridge, Masss.,U.S.A.
 (Drake et al. 2002) $-$ studied independently two objects, 3C58 in Cassiopeia, 
 and RXJ1856.5 $-$ 3754 in the outskirts of the RCrA dark molecular cloud in
 Corona Australis respectively by combining data from NASA's Chandra X-ray 
 Observatory, and the Hubble Space Telescope. These objects, at first, seemed 
 to be ordinary neutron stars. However, when observed more carefully, each of 
 them showed evidence of being an even smaller and denser   object, a quark 
 star or strange star.
 
The team led by David Helfand failed to detect the 
  expected X- radiation  from the hot surface of 3C 58, a putative neutron star,
   believed to be the remnant core of a supernova explosion witnessed by Chinese
    and Japanese astronomers in A.D. 1181. This led the team  to conclude that 
    3C 58 has a surface temperature less than $10^{6}$K, a value far below the predicted 
value, assuming that the object is a neutron star.

The group led by Drake analyzed deep 
  \textit{Chandra} Low Energy Transmission  Grating and High Resolution Camera Spectroscopic 
  observations of the isolated putative neutron  star RX J1856.5 $-$ 3754 with 
  a view to searching for metallic and resonance cyclotron spectral features
 and for pulsation behaviour.  The group found that the X-ray spectrum is well 
 represented by an $\sim$60 eV ( $7\times10^{5}\mathrm{K}$ ) blackbody. It did not find any unequivocal 
 evidence of spectral line or edge features which argues against the metal dominated 
 models. The group also found that the data did not contain any evidence for pulsation.
 The ``radiation radius'' $R_{\infty}$ was found  to be $3.8-8.2 \mathrm{km}$,
  where
\begin{equation}  
R_{\infty} = \frac{R}{(1-2GM/Rc^{2})^{1/2}}
\end{equation} 
R being the true radius of the star of mass M. The group is of the view that the combined
 observational evidence $-$ a lack of spectral and temporal features, and an implied radius 
 $R_{\infty}=3.8-8.2 \mathrm{km}$ that is too small for current neutron star models $-$ points to a more 
 compact object, a quark star, rather than a neutron star. According to Drake et al.(2002),
 of the existing quark star candidates, RX J 1856.5-3754 presents the strongest and the most
  direct case. 


\section{Transition from Hadrons to Quarks}
If indeed the quark stars exist, the pertinent question is $\colon$ How are they formed ?  
The answer to this question lies in understanding the physical process that leads to the
transition  of ordinary matter consisting of hadrons (i.e. baryons and mesons) and leptons
into quark-gluon plasma (QGP) permeated by leptons . In this context it may be noted that
 though actual existence of quarks $-$ up(u), down(d) charm(c), strange(s), top(t), bottom(b) $-$ has
  been only indirectly confirmed by experiments that probe hadronic structure by means of 
  electromagnetic and weak interactions, and by production of various quarkonia ($\bar{q}q$), 
the bound states of quarks (q) and antiquarks ($\bar{q}$), in high energy collisions made
possible by various high energy particle accelerators, no \textit{free} quark has, so far, been
detected in experiments at these accelerators. This fact has been attributed to the phenomenon
called \textit{infrared slavery} of quarks, i.e. to the nature of the interaction between quarks
responsible for their \textit{confinement} inside hadrons. On the contrary, the results of deep
inelastic scattering experiments reveal an altogether different feature of the interaction
between quarks. If one examines quarks at very short distances ($< 10^{-13} \mathrm{cm}$) by
observing the scattering of a nonhadronic probe, e.g., an electron , or a neutrino, one finds
that quarks move almost freely inside baryons and mesons as though they are not bound at all.
This phenomenon is called  \textit{asymptotic freedom} of quarks. In fact, Gross and Wilczek (1973 a,b),
 and Politzer(1973) have shown that the running coupling constant of the interaction between two 
 quarks vanishes in the limit of infinite momentum (or,equivalently in the limit of zero
  separation). Consequently, in order to liberate quarks from \textit{infrared slavery}, i.e. for 
  quark \textit{deconfinement}, very large energy, more than what is available in the existing 
  terrestrial particle accelerators, is required. In fact, it has been shown theoretically that 
  when the energy \textit{E} of the particles $\sim 10 ^{2} \mathrm{GeV}$ ( the separation \textit{s} 
  between the particles $\sim 10 ^{-16} \mathrm{cm}$ ) corresponding to a temperature \textit{T} 
  $\sim 10 ^{15} \mathrm{K}$, all interactions are of the Yang-Mills type with
   $\mathrm{SU_{c}(3)}\times \mathrm{ SU_{I_{W}}(2)}  \times  \mathrm {U_{Y_{W}}(1)}$ 
   gauge symmetry, where \textit{c} stands for colour, \textit{$\mathrm{ I_{W}}$} for weak isospin,
   and \textit{$\mathrm{ Y_{W}}$} for weak hypercharge , and at this stage quarks are liberated 
   from \textit{infrared slavery}, and acquire \textit{asymptotic freedom}, i.e. 
   \textit{quark deconfinement} occurs as a result of which matter now consists of its fundamental 
   constituients $\colon$ spin 1/2 leptons, viz., the electrons, the muons, the tau leptons, and 
   their neutrinos, which interacts only through \textit{electro-weak interaction} (i.e. the 
   unified electromagnetic and weak interactions) ,and the spin 1/2 quarks; u,d,c,s,t,b, which 
   interact \textit{electroweakly} as well as through the \textit{colour force} generated by gluons(G)
   (Ramond 1983). In other words, when \textit{E}$\geq 10^{2} \mathrm{GeV} $
    (\textit{s}$\leq 10^{-16} \mathrm{cm}$) corresponding to  \textit{T}$\geq 10^{15} \mathrm{K} $,
    the entire matter is converted into quark-gluon plasma permeated by leptons.

\section{Experimental Evidence for Existence of QGP}

It may be emphasized here that the notion of QGP is not just a theoretical speculation, or 
conjecture. There are positive indications of its existence as revealed by the series of 
experiments performed at CERN, the European Organization for Nuclear Research at Geneva,
and at RHIC, the Relativistic Heavy Ion Collider, the world's newest and largest particle 
accelerator for nuclear research , at Brookhaven National Laboratory in Upton, New York 
(Heinz 2001). Programmes to create QGP in terrestrial laboratories are already in progress at
CERN and RHIC.

Recently, a team of 350 scientists from 20 countries almost succeeded in creating QGP at CERN
 by smashing together lead ions at temperatures $\sim 10^{12}\mathrm{K}$, and densities 
 $\sim 20$ times that of nuclear matter. A report released by CERN on February 10, 2000 said,  ``
 A series of experiments using CERN's lead beam have provided compelling evidence for the 
 existence of a new state of matter in which quarks, instead of being bound up in more complex 
 particles such as protons and neutrons, are librated to roam freely.''
 
 On the other hand, RHIC's goal is to create QGP by head-on collisions of two beams of gold ions 
 at energies 10 times that of CERN's, and densites 30 times that of nuclear matter, which is 
 expected to produce QGP with higher temperature and longer lifetime, thereby allowing much 
 clearer and direct observation. The programme at RHIC began in the summer of 2000, and after
  two years, Thomas Kirk, Brookhaven's Associate Laboratory  Director for High Energy Nuclear
  Physics, remarked, ``It is too early to say that we have discovered the quark-gluon plasma
  , but not too early to mark the tantalizing hints of its existence.'' 
  
Later, on June 18, 2003 a special scientific colloquium was held
 at Brcokhaven Natioal Laboratory (BNL) to discuss the latest findings at RHIC.
 At the colloquium, it was announced that in the detector system known as STAR ( Solenoidal Tracker AT RHIC ) head-on collision between two beams of gold nuclei of energies of 130 GeV per nuclei resulted in the phenomenon called ``jet quenching``. STAR as well as three other experiments at RHIC viz., PHENIX, BRAHMS, and
 PHOBOS, detected suppression of ``leading particles``, highly 
 energetic individual particles that emerge from nuclear fireballs, in gold-gold collisions. Jet quenching and leading particle suppression are signs
 of QGP formation. The findings of the STAR experiment were presented at the 
 BNL colloquium by Berkeley Laboratory's NSD ( Nuclear Science Division ) physicist Peter Jacobs.   
  
  However, the experimental evidence of the QGP is indirect and leaves much to be done to
   definitively confirm the existence of QGP. In view of this, CERN 
   will start a new experiment ALICE, soon (around 2007 - 2008 at 
   its Large Hadron Collider (LHC) in order to definitively and 
   conclusively create QGP.
   
   Obviously, the lack of complete success in creating QGP in
   terrestrial laboratories is due to the fact that these laboratories fall short of the threshold energy
 required for creating QGP. However, in the universe we have already naturally occurring 
 ultrahigh energy particle accelerators, as will be shown in section 5, in the form of 
 gravitationally collapsing black holes, wherein not only this threshold of energy, but
  even much more can be reached.
  
\section{Internal Dynamics of a Gravitationally Collapsing Black Hole}

Attempts have been made, using the general theory of relativity (GTR), to understand what happens
inside a gravitationally collapsing black hole. In doing so, various simplifying assumptions have 
been made. In the simplest treatment of Oppenheimer and Snyder(1939) a black hole is considered to be
a ball of dust with negligible pressure, uniform density $\rho = \rho(t)$, and at rest at $t=0$.
These assumptions lead to the unique solution of the Einstein field equations, and in the comoving 
co-ordinates the metric inside the black hole is given by

\begin{equation}
ds^{2} = dt^{2} - R^{2}(t) \left[ \frac{dr^2}{1-k r^{2}} + r^{2}d\theta^{2} + r^{2}sin^{2}\theta d\phi^{2}\right] 
\end{equation}        
in units in which c, the speed of light in vacuum, is  1, and where $k$ is a constant (Weinberg
1972a). The requirement of energy conservation implies that $\rho$(t) R$^{3}$(t) remains constant. On normalizing the radial co-ordinate r so as to have $R(0)=1$, on gets $\rho(t)= \rho(0)R^{-3}(t) $.
Furthermore , since the fluid is assumed to be at rest at $t=0$,i.e., $\dot{R}(0)=0$, the field equations give 
$k = 8\pi G \rho(0) /3 $. Finally, the solution of the field equations is given by the parametric
equations of a cycloid $\colon$
\begin{equation} 
t = \left( \frac{\psi+sin\psi}{2\sqrt{k}} \right) 
\end{equation} 
  
\begin{equation}
R = \frac{1}{2}(1+cos\psi)
\end{equation} 
Equation (4) implies that when $\psi=\pi$,i.e. when

\begin{equation} 
t = t_{s} = \frac{\pi}{2\sqrt{k}} = \frac{\pi}{2}\left( \frac{3}{8\pi G \rho(0)}\right) ^{\frac{1}{2}} 
\end{equation} 
a space time singularity occurs; the scale factor R(t) vanishes. In other words, a black hole of uniform 
density having the initial value $\rho(0)$, and zero pressure collapses from rest to a point in 3-space,
 i.e., to a 3-subspace of infinite curvature and zero proper volume, in a finite time $ t_{s} $ ; the collapsed state 
 being a state of infinite proper energy density. 
 
 Actually, the internal dynamics of non-idealized, real black hole
  is very complex. Even in the case of a spherically symmetric collapsing black hole with non zero pressure 
  the details of interior dynamics are not well understood, though major advances in the understanding are 
  being made by means of numerical computations and analytic analyses. But in these computations and analyses 
  no new features have emerged beyond those that occur in the simple uniform-density, free fall
collapse considered above (Misner, Thorne, and Wheeler 1973). However, using topological methods Penrose(1965,1969),
 Hawking(1966a, 1966b, 1967a, 1967b), Hawking and Penrose(1970), and Geroch(1966, 1967, 1968) have proved a number 
of singularity theorems purporting that if an object of mass M contracts to dimensions smaller than the 
gravitational radius $r_{g} = 2 GM/c^{2}$, i.e., if it crosses the event horizon, and if other reasonable 
conditions $-$ namely, validity
 of the GTR, positivity of energy, ubiquity of matter, and causality $-$ are
 satisfied, its collapse to a singularity is inevitable. But the question is $\colon$ Has the validity of the GTR 
 been established experimentally in the case of strong fields.? The answer is $\colon$ Certainly not. Actually, 
 the GTR has been experimentally verified only in the limiting case of weak fields. Moreover, it has been 
 demonstrated theoretically that when curvatures exceed the critical value $C_{g} = 1/{L_{g}}^{4} $, where 
 $L_{g} = (\hbar G/c^{3})^{1/2} = 1.6\times 10^{-33} \mathrm{cm}$  corresponding to the critical density 
 $ \rho _{g} = 5 \times 10^{93} g \mathrm{cm}^{-3}$, the GTR is no longer valid; quantum effects must enter 
 the picture( Zeldovich and Novikov 1971). Therefore, it is clear that the GTR breaks down before a gravitationally
 collapsing  object could collapse to a singularity. Consequently, the conclusion based on the GTR that 
 any gravitationally collapsing object of mass greater than the Oppenheimer-Volkoff limit ($\sim 3 \mathrm{M}_\odot$)
 in general, and a black hole in particular, collapses to a singularity need not be held sacrosanct, actually it may
 not be correct at all. 
 
 It may also be noted that while arriving at the singularity theorems attention has been focused mostly on the 
 space - time geometry and geometrodynamics; matter has been tacitly treated as a classical fluid, remaining entirely 
 unchanged structurally even on being crushed heavily during the gravitational collapse. This is not tenable. To 
 begin with, when the density of matter in a collapsing object reaches the value $\rho \sim 10^{7} \mathrm {g cm^{-3}}$,
the process of neutronization sets in; the electrons and protons in the object combine into  neutrons through the 
 reaction 
 \begin{equation} 
 p + e^{-} = n + \nu _{e} 
\end{equation} 
  The electron neutrinos $ \nu _{e} $ so produced escape from the object. During the gravitational contraction when 
 the density reaches the value $\rho \sim 10^{14} \mathrm {g cm^{-3}}$, the object consists almost entirely of neutrons.
 Of course, enough electrons and protons must remain in the object so that Pauli's exclusion principle prevents neutron
 beta decay 
  \begin{equation}
  n \rightarrow p + e^{-} + \bar{\nu _{e}}
  \end{equation}
  where  $\bar{\nu _{e}}$ is the electron antineutrino (Weinberg 1972b). Therefore, when a black hole collapses to a 
  density $\rho \sim 10^{14} \mathrm {g cm^{-3}}$, it would consist almost entirely of neutrons apart from traces of 
  protons and electrons. However, neutrons as well as 
  protons and electrons are fermions, and as such they obey Pauli's exclusion principle. If a black hole collapses to a 
singularity, i.e., to a point in 3-space, then all the neutrons in the black hole would be crammed into just two quantum 
states available at that point, one for spin up, and another for spin down neutron. This would violate Pauli's exclusion 
principle according to which no more than one fermion of a given species can occupy any quantum state. So would be the 
case with the protons and the electrons in the black hole. Consequently, either Pauli's exclusion principle would be 
violated, or a black hole would not collapse to a singularity in contravention to Pauli's exclusion principle. It may be 
recalled, however, that Pauli's exclusion principle has a profound theoretical basis, it is a consequence of the 
microcausality in local quantum field theory (Huang 1982).In addition to this, it has been experimentally validated
  in the realms of atomic, subatomic, nuclear, and subnuclear physics, both at low energies, and at high and ultrahigh
   energies. On the contrary,this is not the case with the GTR .

\section{Gravitationally Collapsing Black Hole: An Ultrahigh Energy Particle Accelerator}

For creating QGP we need an ultrahigh energy particle accelerator which can accelerate particles to energies \textit{E $\geq 10^{2}$} GeV 
corresponding to temperatures \textit{T $\geq 10^{15} $} K. At present such an accelerator does not exist in 
any terrestrial laboratory. But as mentioned towards 
the end of section 3, the universe has such ultrahigh energy particle accelerators in the form of gravitationally collapsing
black holes.
 
 To see this we consider a gravitationally collapsing black hole. On neglecting mutual interactions
the energy \textit{E} of any one of the particles comprising the black hole is given by\textit{ $E^{2} = p^{2} + m^{2} >p^{2}$}, 
in units in which the speed of light in vaccum, \textit{$c=1$}, and where \textit{$p$} is the magnitude of the 3-momentum of the particle 
, and \textit{m} its mass. But \textit{$ p = h/\lambda$}, where \textit{$\lambda$} is the de Broglie wavelength of the particle,
 and \textit{h} Planck's constant of action. Since all lengths in the collapsing black hole scale down in proportion to the 
 scale factor \textit{R(t)} in equation (2), it is obvious that \textit{$\lambda \propto R(t)$}. Therefore, 
 \textit{$p \propto R^{-1}(t)$}, and hence \textit{$ p = a R^{-1}(t)$}, where \textit{$a$} is the
constant of proportionality . This implies that \textit{$E > a / R$}. Consequently, \textit{$E$} as well as \textit{$p$} 
increases continually as \textit{$R$} decreases. It is also obvious that \textit{$E$} and \textit{$p\rightarrow \infty$}
as \textit{$R\rightarrow 0$}. Thus, in effect, we have an ultrahigh
energy particle accelerator in the form of a gravitationally collapsing black hole, which can, in the
absence of any physical process inhibiting the collapse of the black hole to a singularity, accelerate
particles to an arbitrarily high energy and momentum without any limit.
 
 What has been concluded above
can also be demonstrated alternatively, without resorting to the GTR as follows. As an object collapses
under its selfgravitation, the interparticle distance \textit{$s$} between any pair of particles in the object
decreases, Obviously, the de Broglie wavelength \textit{$\lambda$} of any particle in the object is less than, or equal
to \textit{$s$}, a simple consequence of Heisenberg's uncertainty principle. Therefore, \textit{$ s \geq h / p$}, 
where \textit{$h$} is Plank constant of action, and \textit{$p$} the magnitude of the 3-momentum of the particle, 
Consequently, \textit{$p \geq h/s$}, and hence \textit{$ E \geq h/s $}. Since during the collapse of the object \textit{s} decreases,
the energy \textit{$E$} as well as the momentum \textit{$p$} of each of the particles in the object increases. 
Moreover, from \textit{$E \geq h/s$} and \textit{$ p \geq h/s$} it follows 
that \textit{$E$} and \textit{$p \rightarrow \infty $} as \textit{$s \rightarrow 0$}. Thus, any gravitationally collapsing 
object in general, and a black hole in particular, acts as an ultrahigh energy particle accelerator.
 
 It is also obvious 
that  \textit{$\rho$}, the density of matter in the black hole, increases as it collapses. In fact,
 \textit{$\rho \propto R^{-3}$}, and hence $\rho \rightarrow \infty$  as  $R \rightarrow 0$.

\section{The End-Point and the End-product of a Black Hole}

If indeed a black hole collapses  to a singularity, then the most pertinent question is: What happens to
a black hole after it collapses to a singularity, i.e., to a point in 3-space, to a state of infinite proper
energy density (e.g., after \textit{t} $>$ \textit{$t_{s}$} in the Oppenheimer - Synder black hole) ? 
Will the particles of infinite energy and momentum remain frozen at a point forever 
after it collapses to a singularity ? Certainly not, it is inconceivable how particles 
of infinite energy and momentum would remain frozen  at a point forever. Instead, some 
thing would happen before the collapse of a black hole to a singularity that would 
 avert its collapse to a singularity. 
 
 Besides, no interaction other than the gravitational interaction has been 
taken into account while arriving at the singularity
theorems, it has been tacity assumed that the other interactions are negligible. But
this tacit assumption is not justified; for, at ultrahigh energies and ultrahigh
 densities quantum chromodynamics plays an importent role as discussed in
 section 2. As shown in section 5, the energy \textit{$E$} of the particles comprising the
matter in a gravitationally collapsing black hole continually increases, and
so does the density $\rho$ of the matter in the black hole whereas the separation
\textit{$s$} between any pair of particles decreases. Consequently, during the final stages
of the gravitational collapse of a black hole the following scenario may occur$\colon$

When \textit{$ E \geq 10^{2} $ }GeV corresponding to the temperature $T \geq 10^{15}$ K, 
and \textit{$s \leq 10^{-16}$} cm, the entire matter in the black hole would be converted 
into QGP permeated by leptons. Even after this the gravitational collapse of the black hole 
may continue, but only upto the stage
when the ` gas '  ( i.e., the ensemble ) of each and every species of quarks and leptons becomes
fully degenerate, i.e., when all the quantum states upto the relevant Fermi surface in
the momentum space are fully occupied. The collapse of a black hole to a singularity  would
be inhibited by Pauli's exclusion principle. For, if a black hole collapses 
to a singularity, i.e. to a
point in 3-space, then all the quarks and leptons of each species would be crammed
 into just two quantum states available at that point for the quark or lepton of
that species, one for spin up, and another for spin down quark or lepton of that
species.This would violate Pauli's exclusion principle, because quarks and
leptons are fermions, being spin $1/2$ particles, and as such they are governed
 by Pauli's exclusion principle according to which no more than one fermion of 
  a given species can occupy any quantum state available for that species.
  
 If a black hole cannot collapse to a singularity in contravention to Pauli's
 exclusion principle, then the most pertinent question is : What will happen to
 it eventually ? Eventually, a black hole may explode, a mini bang of a sort
 may occur, and after explosion, much of the matter in it may expand even 
 beyond the event horizon, and the remnant core may stabilize as a quarks star as it already consists of quarks,
  gluons, and leptons. This may happen, notwithstanding Zeldovich and Novikov's
 (1971) strong assertion that after a collapsing sphere's radius decreases to 
 $r < r_{g}$ in a finite proper time its expansion into the external space from 
 which the the contraction originated is impossible, even if the passage of 
 matter through infinite density is assumed.
 
 A gravitationally collapsing black hole may also explode by the very same
 mechanism by which the big bang occurred in the universe, if indeed it did 
 occur. This can be seen as follows. At the present epoch the volume of the 
 universe is $ \sim 1.5 \times 10^{85} cm^{3} $ and the density of the galactic material 
 through out the universe is $ \sim 2 \times 10 ^{-31} g cm^{-3} $ (Allen 1973). Hence a 
 conservative estimate of the mass of the universe is 
 $ \sim 1.5 \times 10^{85} \times 2 \times 10^{-31} g = 3 \times 10^{54} g$ (actually it would be much
 more if the mass of the intergalactic matter as well as that of the dark 
 matter in the universe is taken in to account). However, according to Gamow's big 
 bang model, before the big bang, the entire matter in the universe was 
 contained in the \textit{ylem} which occupied very very small volume. The gravitational
 radius of the ylem of mass $3 \times 10^{54} g$ was $4.45 \times 10^{21} km$ (it would be 
 much larger if the actual mass of the universe were taken into account which 
 is greater than $3 \times 10^{54} g$ ). Obviously, the radius of the ylem was many 
 order of magnitude smaller than its gravitational radius, and yet the ylem 
 exploded with a big bang, and in due course of time, its expanding matter 
 crossed the event horizon and expanded beyond it up to the present Hubble 
 distance $c/H_{0} \sim 1.5 \times 10^{23} km$ where $c$ is the speed of light in vacuum
 and $H_{0}$ the Hubble constant at the present epoch. Consequently , if the 
 ylem could explode and its matter could cross the event horizon and expand 
 beyond it in spite of Zeldovich and Novikov's assertion to the contrary, why
 can't a gravitationally collapsing black hole also explode, and much of the 
 matter in in it expand beyond the event horizon in due course of time ? 
 However, the mechanism by which the ylem exploded is not definitively known and 
 as such the mechanism by which a black hole would explode before collapsing to
 purported singularity is also not known.
 
 Another way of looking at the problem is the following. It may not be 
 unreasonable to assume that, during the gravitational collapse, the outward 
 pressure $P$ inside a gravitationally collapsing black hole increases 
 monotonically with the increase in the density of matter, $\rho$. Actually, it 
 may be given by the polytrope, $P = K \rho^{\frac{(n+1)}{n}} $, where $K$ is a constant 
 and n is the polytropic index. Consequently, $P \rightarrow \infty$ as $\rho \rightarrow \infty$, i.e. $P \rightarrow \infty$ as the scale factor $ R(t) \rightarrow 0 $ ( or,
 equivalently $ s \rightarrow 0 $ ). In view of this, during the gravitational 
 collapse of a black hole, at a certain stage when the density of matter $\rho = \rho_{c}$, the 
 outward pressure $P = K \rho_{c}^{\frac{(n+1)}{n}}$, inside the black hole may be large 
 enough to withstand the inward gravitational force, and the object may become
 gravitationally stable and thus end up as a stable quark star since it 
 consists of quarks, gluons and leptons.

 This scenario also explains the absence of a large number
of black holes in the universe.In principle, like white dwarfs and neutron stars, there
 should be quite a large number of black holes in every galaxy. White dwarfs
and neutron stars are the end products in the sequence of evolution of stars with
mass less than the Chandrasekhar and the Oppenheimer-Volkoff limits
 respectively whereas black holes are the end products in the sequence
 of evolution of stars with mass more than the Oppenheimer-Volkoff limit.
  Therefore, there should not be an inequable distribution of black holes
in the universe in general, and in any galaxy in particular.But contrary
 to this expectation, very few black holes have been observed.Though black
 holes cannot be observed directly, they would manifest their presence 
by the strong gravitational fields produced by them in their vicinity which would
bend the ray of light appreciably, and perturb the motion of the
celestial bodies passing by, or in its vicinity, but outside
 the event horizon. 
 
 Of course, an alternative possibility is
 the generally held view that a black hole finally collapses
 to a singularity.If this is true, then this presents an
evidence of the violation of Pauli's exclusion principle.

\section{Conclusion}
 
A gravitationally collapsing black hole acts as an ultrahigh energy particle accelerator
 that can accelerate particles comprising the matter in the black hole to inconceivably
  high energies.During the continual acceleration of particles, as a result of the continual
 gravitational collapse of the black hole, a stage will be reached when the energy of the 
   particles $E\sim10^{2}$ GeV corresponding to the temperature $T\sim10^{15}$ K. At this 
   stage the entire matter in the black hole will be converted into quark - gluon plasma 
   permeated by leptons. With further collapse one of the two possibilities will occur, 
   either Pauli's exclusion principle would be violated and the black hole would eventually 
   collapse to a singularity, or Pauli's exclusion principle would hold good and would avert 
   the collapse of the black hole to a singularity, and eventually the black hole would explode 
   with a mini bang of a sort. Finally, the remnant core would stablize as a quark star.\\

\begin{acknowledgements}
 The author thanks Professor S. K. Pandey, Co - ordinator, IUCAA Reference Centre, 
 School of Studies in Physics, Pt. Ravishankar Shukla University, Raipur,
 for making available the facilities of the Centre. He also thanks 
 Laxmikant Chaware and Samriddhi Kulkarni for typing the manuscript.
\end{acknowledgements}

\end{document}